## Original Paper

Zubair Shah[1], Didi Surian[1], Amalie Dyda[1], Enrico Coiera[1], Kenneth D. Mandl[2,3], Adam G. Dunn[1,2]

1. Centre for Health Informatics, Australian Institute of Health Innovation, Macquarie University, Sydney, Australia.
2. Computational Health Informatics Program, Boston Children's Hospital, Boston, United States.
3. Department of Biomedical Informatics, Harvard Medical School, Boston, United States.

# Automatically applying a credibility appraisal tool to track vaccine-related communications shared on social media

## Abstract

**Background:** Tools used to appraise the credibility of health information are time-consuming to apply and require context-specific expertise, limiting their use for quickly identifying and mitigating the spread of misinformation as it emerges. Our aim was to estimate the proportion of vaccine-related Twitter posts linked to webpages of low credibility and measure the potential reach of those posts.

**Methods:** Sampling from 144,878 vaccine-related webpages shared on Twitter between January 2017 and March 2018, we used a seven-point checklist adapted from validated tools and guidelines to manually appraise the credibility of 474 webpages. These were used to train several classifiers (random forest, support vector machines, and a recurrent neural network with transfer learning), using the text from a web page to predict whether the information satisfies each of the seven criteria. Estimating the credibility of all other webpages, we used the follower network to estimate potential exposures relative to a credibility score defined by the seven-point checklist.

**Results:** The best performing classifiers were able to distinguish between low, medium, and high credibility with an accuracy of 78%, and labelled low credibility webpages with a precision of over 96%. Across the 144,878 web pages, 14.4% of relevant posts to text-based webpages were of low credibility and made up 9.2% of potential exposures. The 100 most popular links to webpages of low credibility were each potentially seen by between 2 million and 80 million Twitter users globally.

**Conclusions:** The results indicate that while a small minority of low credibility webpages reach a large audience, low credibility webpages tend to reach fewer users than other webpages overall, and are more commonly shared within certain sub-populations. An automatic credibility appraisal tool may be useful for finding communities of users at higher risk of exposure to low credibility vaccine communications.

# Introduction

The spread of misinformation—which we define here to include communications that are not a fair representation of available evidence or communicate that evidence poorly—has become an increasingly studied topic in various domains (1-8). Misinformation can cause harm by influencing attitudes and beliefs (9, 10). While the rapid growth of online communications has benefited public health by providing access to much a broader range of health information, most people trust online health information without attempting to validate the sources (11-13), despite concerns about the presence of misinformation in what they access (14), and known issues where biases and marketing can lead to the miscommunication of evidence (15-19). Proposed approaches for mitigating the impact of misinformation include empowering individuals to better deal with the information they encounter, and improvements in the automatic detection of misinformation in online platforms (1).

Most studies aimed at finding or tracking misinformation on social media do so using a definition based on *veracity*—whether a claim is true or false; real or fake. In the health domain, veracity alone often does not provide enough information to be useful in understanding the range of factors that might influence attitudes and behaviours, such as persuasiveness, timeliness, or applicability. The broader set of factors affecting the quality of how health information is communicated can be described as *credibility* (20). It is important to consider credibility when evaluating the potential impact of health communications on health attitudes and outcomes because certain types of communication can be true but misleading, such as in the case of false balance in news media (21).

A range of tools have been developed to assess the credibility of online health information. Most were designed as checklists to be used by experts to assess the credibility and transparency of what they are reading. The DISCERN tool was designed as a general-purpose tool for evaluating the quality of health information (22), with an emphasis on web pages that patients might use to support the decisions they make about their health. The Quality Index for health-related Media Reports (QIMR) is a more recent example; and differs from previous tools in that it was designed to be used to evaluate the quality of communications about new biomedical research (23). Common elements of the tools used by experts to assess the credibility of health research reporting and patient information online include: the veracity of the included information; transparency about sources of evidence; disclosure of advertising; simplicity and readability of the language; and use of balanced language that does not distort or sensationalise (20). Most of the tools can be time-consuming to use and often require specific training or expertise to apply. Organisations like HealthNewsReview.org, which ended in 2018, use experts to evaluate new health-related communications as they appear in the news media (24).

Public perception of vaccines is an exemplar of the problem of misinformation spread through news and social media (25). Beyond public health and vaccines, previous studies using social media data derived from Twitter to understand the spread and impact of misinformation have variously extracted text from what users post or information about their social connections (26-30). Vaccine attitudes and opinions about disease outbreaks are a common application domain studied in social media research (31-35). In particular, studies of human papillomavirus (HPV) vaccines have made use of the information users post and their social connections, as well as what people might have been exposed to from their networks (36-39). The ability to measure how people engage and share misinformation on social media may help us better target and monitor the impact of public health interventions (40-42).

Given the rate at which new information is made available and the resources needed to appraise them, there is currently no way to keep up with new health-related stories as soon as they appear. While the challenge of managing information volume versus quality was discussed two decades ago (43), methods for managing emerging misinformation in health-related news and media remain an unresolved issue in public health.

### Research objectives

Our aim was to characterise the sharing and potential reach of vaccine-related webpages shared on Twitter, relative to credibility. Because it would not have been feasible to manually assess the credibility of all webpages, we developed and evaluated classifiers to automatically estimate their credibility

## Methods

The study used a retrospective observational design. To estimate the credibility of vaccine-related webpages shared on Twitter, we collected text from vaccination-related webpages by monitoring links from tweets that mentioned relevant keywords. We manually appraised the credibility of a sample of webpages by applying a checklist-based appraisal tool, using the sample to train classifiers to predict a credibility score in unseen webpages. Applying an ensemble classifier to the full set of webpages collected as part of the surveillance, we examined patterns of sharing relative to credibility scores.

## Datasets

We collected 6,591,566 English-language, vaccine-related tweets (including retweets) from 1,860,662 unique Twitter users between 17 January 2017 and 14 March 2018 using the Twitter Search Application Programming Interface (API), using a set of pre-defined search terms (including "vaccin*", "immunis*", "vax*", "antivax*"). For all unique users posting vaccine-related tweets during the study period, we collected the lists of their followers to construct the social network.

We then extracted 1.27 million unique URLs from the set of tweets to identify the set of text-based webpages to include in the analysis. To restrict the set of webpages to only English language text we used a Google library (44); removed webpages with fewer than 300 words in contiguous blocks; and webpages that were internal Twitter links, broken links, or to webpages that are were no longer available. We additionally removed webpages for which most or all of the text was a duplication of other webpages already included, retaining the webpage with the greatest number of words. The remaining set of 144,878 webpages (Figure 1) was used in the subsequent analysis.

To modify how we sampled tweets for constructing a manually labelled dataset, we then used PubMed to search for vaccine-related research articles using search terms "vaccine" or "immunisation" in the title or abstract, automatically expanded by PubMed to include synonyms and MeSH terms. The search returned 306,886 articles. We then used the PubMed IDs of these articles with Altmetric to identify webpages (news, blogs, social media posts) that linked to these articles via their Digital Object Identifier (DOI), PubMed entry, or journal webpage. We found 647,879 unique URLs from Altmetric that cited the selected vaccines-related PubMed articles.

The intersection of the URLs extracted from Altmetric and the URLs extracted from the tweets allowed us to over-sample from the set of webpages for which we expected to have higher credibility scores (described below). This approach also allowed us to exclude most of the URLs shared on Twitter that linked directly to research articles by removing the tweets that were identified by Altmetric.

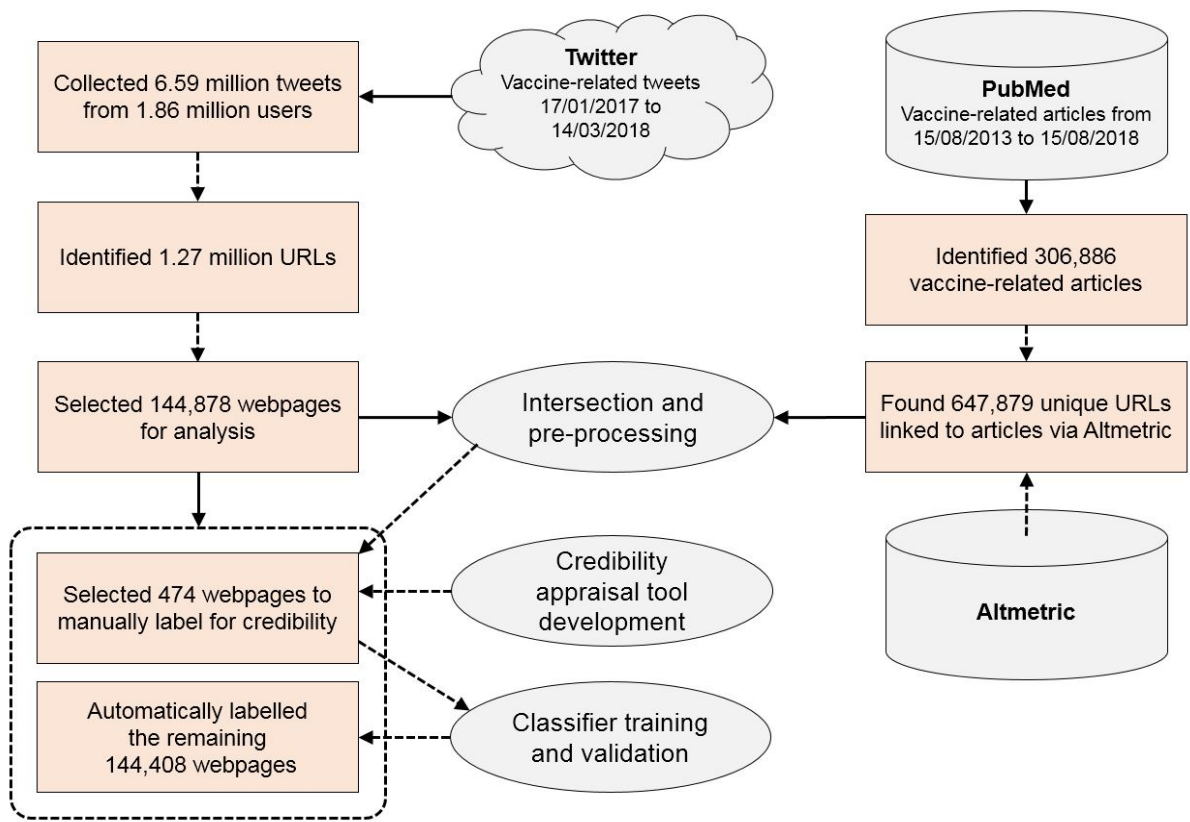

**Figure 1:** The steps used to define the training dataset and automatically label webpages.

We briefly examined the common domains of other webpages that were not amenable to text extraction to better understand the nature of the webpages for which we were not able to estimate a credibility score. Among the set of 1,131,241 URLs that were not included the analysis, 152,553 (13.55%) were links to YouTube and 14,243 (1.26%) were links to Instagram.

## Credibility appraisal tool

The credibility appraisal tool was developed by three investigators (AGD, AD, MS) with expertise in public health, public health informatics, science communication, and journalism. To develop a tool that would work specifically with vaccine-related web pages, the investigators adapted and synthesised individual criteria from a set of checklist based tools and guidelines (20).

- Centers for Disease Control and Prevention guide for creating health materials (45).
- The DISCERN tool (22).
- Health News Review criteria (24), which is informed by Moynihan et al. (46) and the Statement of Principles of the Association of Health Care Journalists (47).
- Media Doctor review criteria (48).
- World Health Organisation report on vaccination and trust (49).
- The Quality Index for health-related Media Reports (QIMR) (23).

Using these documents as a guide, we adapted from the DISCERN and QIMR checklists, and added two additional criteria that were specific to vaccine-related communications. The tool was pilot tested on 30 randomly selected webpages and iteratively refined through discussion among the 3 investigators. The resulting credibility appraisal tool included the following 7 criteria:

(i) information presented is based on objective, scientific research;

(ii) adequate detail about the level of evidence offered by the research is included;

(iii) uncertainties and limitations in the research in focus are described;

(iv) the information does not exaggerate, overstate or misrepresent available evidence;

(v) provides context for the research in focus;

(vi) uses clear, non-technical language that is easy to understand; and

(vii) is transparent about sponsorship and funding.

## Manually labelled sample

The 3 investigators then applied the credibility appraisal tool to an additional 474 vaccine-related webpages. For each webpage, investigators navigated to the website, read the article, and decided whether it satisfied each of the seven criteria. This process produced a set of values (0 or 1) for each criterion and for each webpage. We then summarised the information as a *credibility score*, defined by the number of criteria that were satisfied, and grouped web pages by credibility score into low (from 0 to 2 criteria satisfied), medium (from 3 to 4 criteria satisfied), and high (from 5 to 7 criteria satisfied). Across the 474 expert-labelled examples, the proportion of the web pages that were judged to have satisfied each of the seven credibility criteria varied substantially (Figure 2).

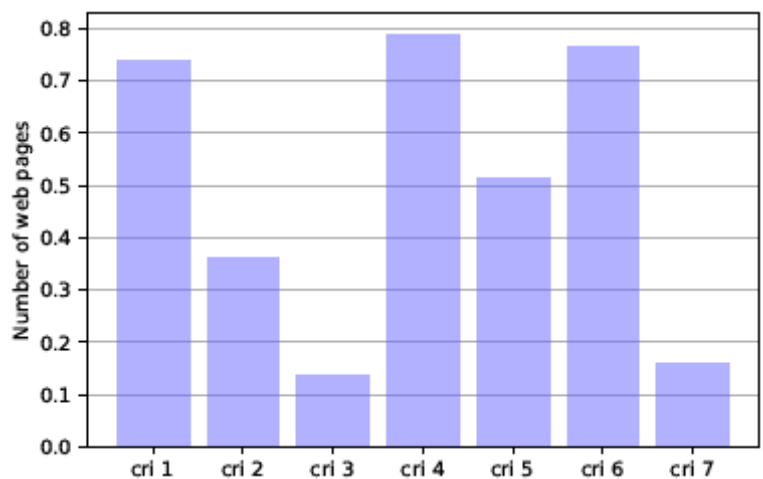

**Figure 2:** The proportion of web pages that met the individual criteria in the 474 web pages used to train the classifiers [cri: criterion].

The investigators independently undertook duplicate appraisals of a subset of the webpages to measure inter-rater reliability, and it was found to be reasonable for separating webpages as low, medium, or high credibility (Fleiss' kappa 0.46; 95% CI 0.41-0.52; p<0.001), and near-perfect when the aim was to separate low credibility webpages from all others (Fleiss' kappa 0.89; 95% CI 0.82-0.97; p<0.001). The design of the checklist suggests that it is a useful approach for identifying webpages of low credibility.

## Classifier design

We compared 3 machine learning methods that are commonly used for document classification problems: support vector machines (SVM), random forests (RF) and recurrent neural networks (RNN). The SVM approach trains a large-margin classifier that aims to find a decision boundary between two classes that is maximally far from any point in the training data. Classifiers that make us of the RF approach construct sets of classification trees by randomly selecting a subspace of features at each node of the decision tree to grow its branches. It then uses bagging to generate subsets of training data for constructing individual trees, which are then combined to form random forests model. The RNN approach refers to a class of artificial neural networks comprised of neural network blocks that are linked to each other to form a directed graph along a sequence. The method is used to model dynamic temporal behaviour for a time sequence, which is useful for understanding the language. These classifiers are shown to outperform in text classification problems in a variety of domains; achieving state-of-the-art results on standard academic benchmark problems (50).

The aim of these supervised machine learning techniques was to train a model to predict the class of an unseen document by learning how to distinguish the language used across classes. To apply the classifiers, we cleaned the text downloaded from web pages by removing extra spaces, tabs, extra newlines, and non-standard characters including emoticons. Each web page was then included as a document in our corpus.

To develop the RNN classifier, we used average stochastic gradient descent (ASGD) weight dropped (AWD) long short-term memory (LSTM) (51). In what follows, we refer to this as the deep learning (DL) based classifier. The DL-based classifier comprised a backbone and a custom head. The backbone is a language model which is a deep recurrent neural network. The head is a linear classifier comprising two linear blocks with ReLU activations for the intermediate layer and a softmax activation for the final layer that can estimate the target labels (in our case, whether it satisfies a credibility criterion).

Language models are trained to understand the structure of the language used in a corpus of documents, and its performance is measured by its ability to predict the next word in a sentence based on the set of previous words. After the language model is trained for this task, the complete DL-based classifier is then fine-tuned to predict whether a document satisfies each of the credibility checklist criteria. Language models are often trained to learn the structure of the language in a target corpus, but recent advances in transfer learning have produced superior results including shorter training times and higher performance. An example is the Universal Language Model Fine Tuning (ULMFiT) method (50), which was proposed and evaluated on NLP tasks.

We used transfer learning to create the language model backbone. The language model was developed with 3 layers, 1,150 hidden units and an embedding size of 400 per word, and the weights were initialised from a pre-trained wikitext103 language model produced by Howard et al. (50). For the DL-based classifier we report in the results below. The parameters and values used in the initialisation of the language model and classifier are given in Table 1. The results of the performance of the associated language model are given in Figure 3.

**Table 1:** The parameters and corresponding values for the initialisation of the language model and classifier

| Parameters | Value |
|---|---|
| **Weight decay** | 1.00E-04 |
| **BPTT** | 60 |
| **Batch size** | 52 |
| **Drop outs** | [0.25, 0.1, 0.2, 0.02, 0.15] |
| **Embedding size** | 400 |
| **Number of layers** | 3 (language model), 5 (classifier) |
| **Optimiser** | Adam [27] |
| **$\beta_1$, $\beta_2$** | 0.8, 0.99 |

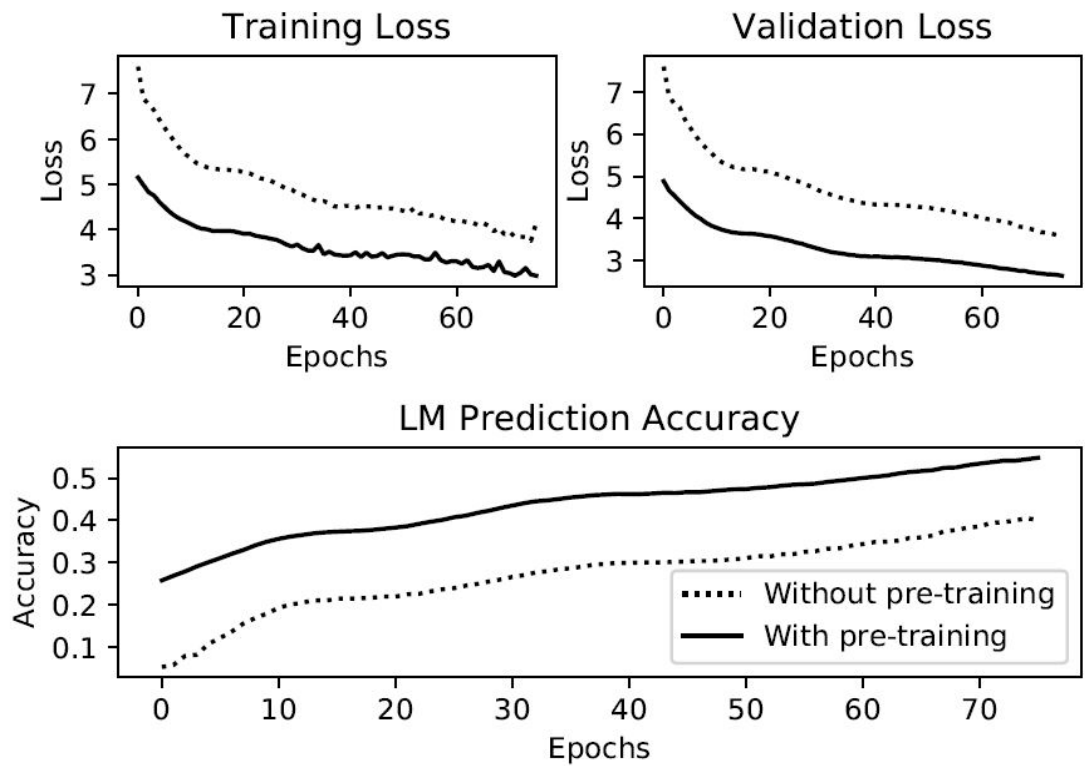

**Figure 3:** The performance difference of the language model (LM) for two different settings, including training loss (top-left), validation cross-entropy loss (top-right), and the accuracy of the language model predicting the next word in a sentence given previous words, in the validation text (bottom).

For the SVM and RF based classifiers, we performed additional pre-processing to remove stop-words and low-frequency words to improve accuracy. After pre-processing, there were 60,660 unique words used across the entire corpus; these were used as features for training and testing RF and SVM classifiers. Each document was represented as a set of feature vectors, where features were defined by term frequency, inverse document frequency (TF-IDF) weights. TF-IDF represents the importance of a word to a document in a corpus, which increases proportionally to the number of times it appears in the document but is offset by the frequency of the word in the corpus, ensuring the similarity between documents be more influenced by discriminative words with relatively low frequencies in the corpus. The best parameters for SVM and RF are found using grid search functionality of *scikit-learn* library and are given in Table 2.

**Table 2:** The parameters used for SVM and RF classifiers, all other parameters are kept as default.

| Parameters | Value |
|---|---|
| **SVM** | |
| C | 100 |
| gamma | 1 |
| kernel | linear |
| norm | l1 |
| use-idf | TRUE |
| max-df | 1 |
| ngram-range | (1, 1) |
| **RF** | |
| n-estimators (RF) | 10 |
| criterion (RF) | gini |
| min-impurity-split (RF) | 1.00E-07 |

Using the expert-labelled data we trained 21 classifiers (one per criterion for each of the RF, SVM and DL-based classifiers) and evaluated the performance of the classifiers in 10-fold cross-validation tests, reporting the average $F_1$-score and accuracy for all three classifiers. While the comparison of the performance across the set of classifiers may be of interest, our aim was to provide the basis for an ensemble classifier that could reliably estimate which of criteria met by each webpage.

### Sharing and potential exposure estimation

Following the development of a reliable tool for automatically estimating the credibility of vaccine-related communications at scale, we aimed to characterise patterns of potential exposure to low credibility vaccine communications on Twitter. For each web page that met our study inclusion criteria, we estimated its credibility score using the best-performing classifiers for each criterion. We then aggregated the total number of tweets posted during the study period that included a link to the webpage, including tweets and retweets. We then estimated the *potential exposure* by summing the total number of followers for all tweets and retweets. Note that this represents the maximum possible audience and we did not capture the union of individual users who may have been followers of at least one of the users posting the tweet as has been done in previous studies [14].

To examine how users posting links to low credibility webpages might be concentrated within or across sub-populations, we also estimated a per-user measure of credibility, which is defined by the list of credibility scores for any user sharing links to one or more webpages. We used these lists in conjunction with information about followers to construct a *follower network*, which allowed us to identify sub-populations of Twitter users for which the sharing of low credibility vaccine communications was common.

## Results

The RF classifiers produced the highest performance overall, and in most cases predicted whether the text on a vaccine-related web page satisfied each of the credibility criteria with over 90% accuracy (Table 3). The SVM-based classifier produced the highest $F_1$-scores for two of the most unbalanced criteria. Further experiments are needed to determine whether the DL-based classifier outperforms baseline methods if more expert-labelled data are made available. The results show that it is feasible to estimate credibility appraisal for web pages about vaccination without additional human input, suggesting the performance—although variable—is high enough to warrant their use in surveillance.

**Table 3:** Performance of the classifiers (average F1 Score and accuracy in 10-fold cross validation)

|  | DL | | SVM | | RF | |
| --- | --- | --- | --- | --- | --- | --- |
|  | F1 Score | Accuracy | F1 Score | Accuracy | F1 Score | Accuracy |
| **Criterion 1** | 0.851 ±0.005 | 0.740 ±0.008 | 0.903 ±0.032 | 0.842 ±0.045 | **0.950** ±0.015 | 0.924 ±0.019 |
| **Criterion 2** | 0.000 ±0.000 | 0.638 ±0.003 | 0.802 ±0.044 | 0.828 ±0.018 | **0.915** ±0.005 | 0.943 ±0.006 |
| **Criterion 3** | 0.000 ±0.000 | 0.865 ±0.009 | **0.761** ±0.038 | 0.917 ±0.011 | 0.745 ±0.088 | 0.944 ±0.018 |
| **Criterion 4** | 0.882 ±0.001 | 0.789 ±0.002 | 0.903 ±0.042 | 0.833 ±0.068 | **0.959** ±0.017 | 0.936 ±0.022 |
| **Criterion 5** | 0.551 ±0.249 | 0.486 ±0.051 | 0.787 ±0.034 | 0.721 ±0.051 | **0.921** ±0.022 | 0.920 ±0.020 |
| **Criterion 6** | 0.867 ±0.002 | 0.765 ±0.004 | 0.912 ±0.006 | 0.852 ±0.010 | **0.964** ±0.002 | 0.943 ±0.004 |
| **Criterion 7** | 0.000 ±0.000 | 0.840 ±0.008 | **0.801** ±0.029 | 0.924 ±0.006 | 0.764 ±0.057 | 0.936 ±0.004 |

Where the best-performing classifiers were combined to distinguish between low, medium, and high credibility web pages, the overall accuracy of the ensemble classifier that combines best performing classifiers (SVM for criterion 3 and 7 and RF for all other criteria) is 78.30%. In terms of labelling low credibility webpages, the ensemble classifier rarely mislabelled a high or medium credibility webpage as low credibility; more than 96% of the webpages labelled as low credibility were correct.

To consider the expected robustness of the classifiers we additionally analysed the set of terms that were most informative of low credibility webpages. We used a Fisher's exact test to compare the proportion of low credibility webpages a term appeared in at least once relative to the proportion of other webpages in which the term appeared at least once, examining the terms that were over-represented in either direction (Figure 4). The results indicate a set of mostly general terms; terms that are most indicative of low credibility webpages are related to anecdotes and stories, and terms that are most indicative of other webpages are related to research. The results suggest that the sample of webpages used to construct the training data are a broad enough sample to capture general patterns rather than specific differences that would limit the external validity of the approach.

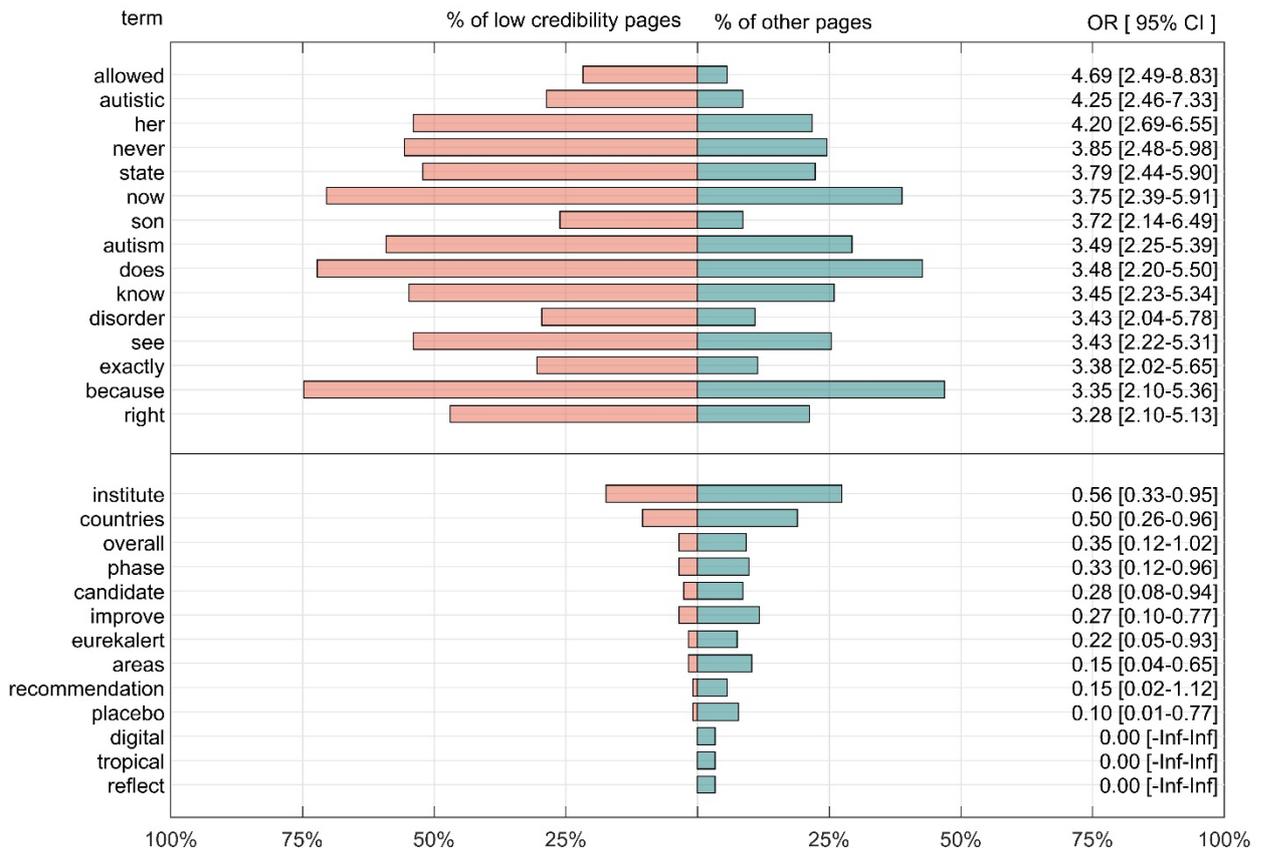

**Figure 4:** A subset of the terms that were informative of low credibility scores in the training set of 474 webpages. Terms at the top are those most over-represented in low credibility webpages, and terms at the bottom are those most over-represented in other webpages [OR: Odds Ratio; CI: Confidence Interval].

Satisfied with the performance of the ensemble classifier, we then applied it to the full set of 144,878 vaccine-related webpages, producing an estimated credibility score for every page. Fewer unique webpages with low credibility scores were shared on Twitter relative to those with medium or high credibility scores (Figure 5), though it is important to consider the performance limitations of the ensemble classifier when interpreting these findings. Measured by the number of times a webpage link was posted on Twitter (including retweets), we found that 14.4% of vaccine-related tweets with links to external webpages were to webpages of low credibility.

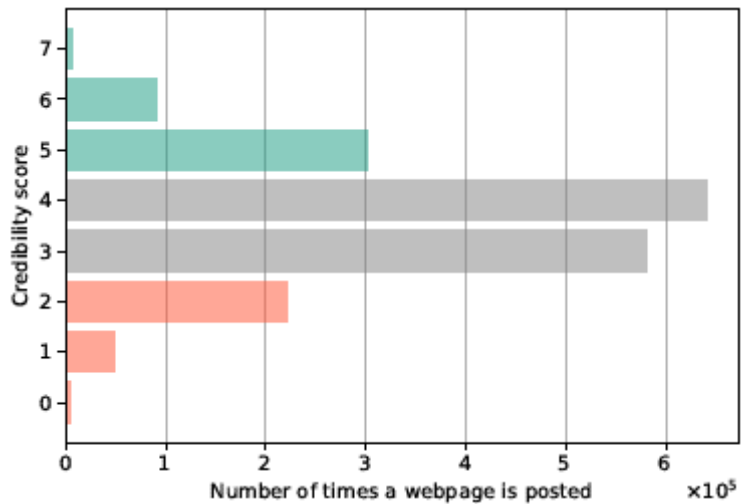

**Figure 5:** The sum of tweets and retweets for links to included web pages relative to the number of credibility criteria satisfied.

When we examined the total number of potential exposures by counting followers, we found that the distribution of total potential exposures per webpage were similar (illustrated by the slope of the three distributions in Figure 6). Measured by the total proportion of exposures to links to relevant webpages, 9.2% of total exposures were to low credibility webpages, and 24.4% of total exposures were to high credibility webpages. Twitter users sharing links to high and medium credibility vaccine-related webpages tended to have a greater number of followers than Twitter users sharing links to low credibility vaccine-related webpages. However, the shape of the distribution shows that some of the low credibility webpages may have been influential; the top 100 by exposure may have been seen by between 2 million and 80 million, and more than 200 had at least 1 million potential exposures.

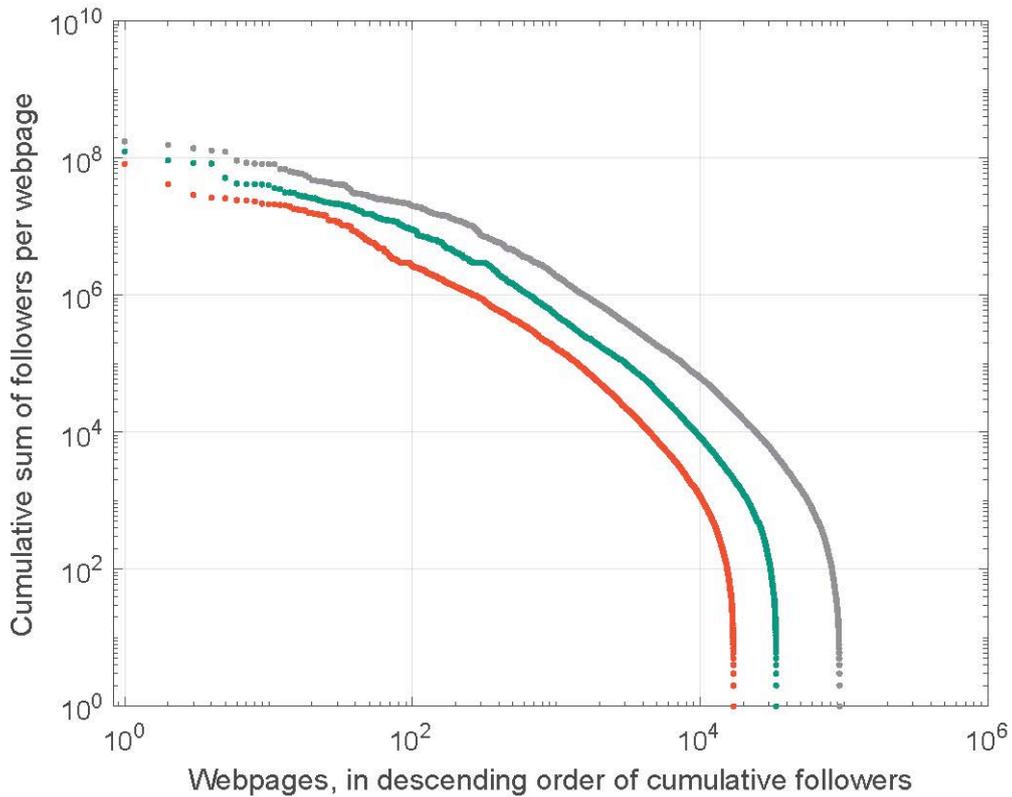

**Figure 6:** The distribution of potential exposures per web page for low (orange), medium (grey) and high (cyan) credibility scores, where low credibility includes scores from 0 to 2, and high credibility includes scores from 5 to 7.

Links to low credibility vaccine-related webpages were more heavily concentrated among certain groups of users posting tweets about vaccines on Twitter. This is evident in a visualization of the follower network for the set of 98,663 Twitter users who posted at least 2 links to webpages included in the study (Figure 7). The network indicates heterogeneity in the sharing of links to low credibility vaccine-related webpages, suggesting that there are likely to be communities of social media users for whom the majority of what they see and read about vaccines is of low credibility.

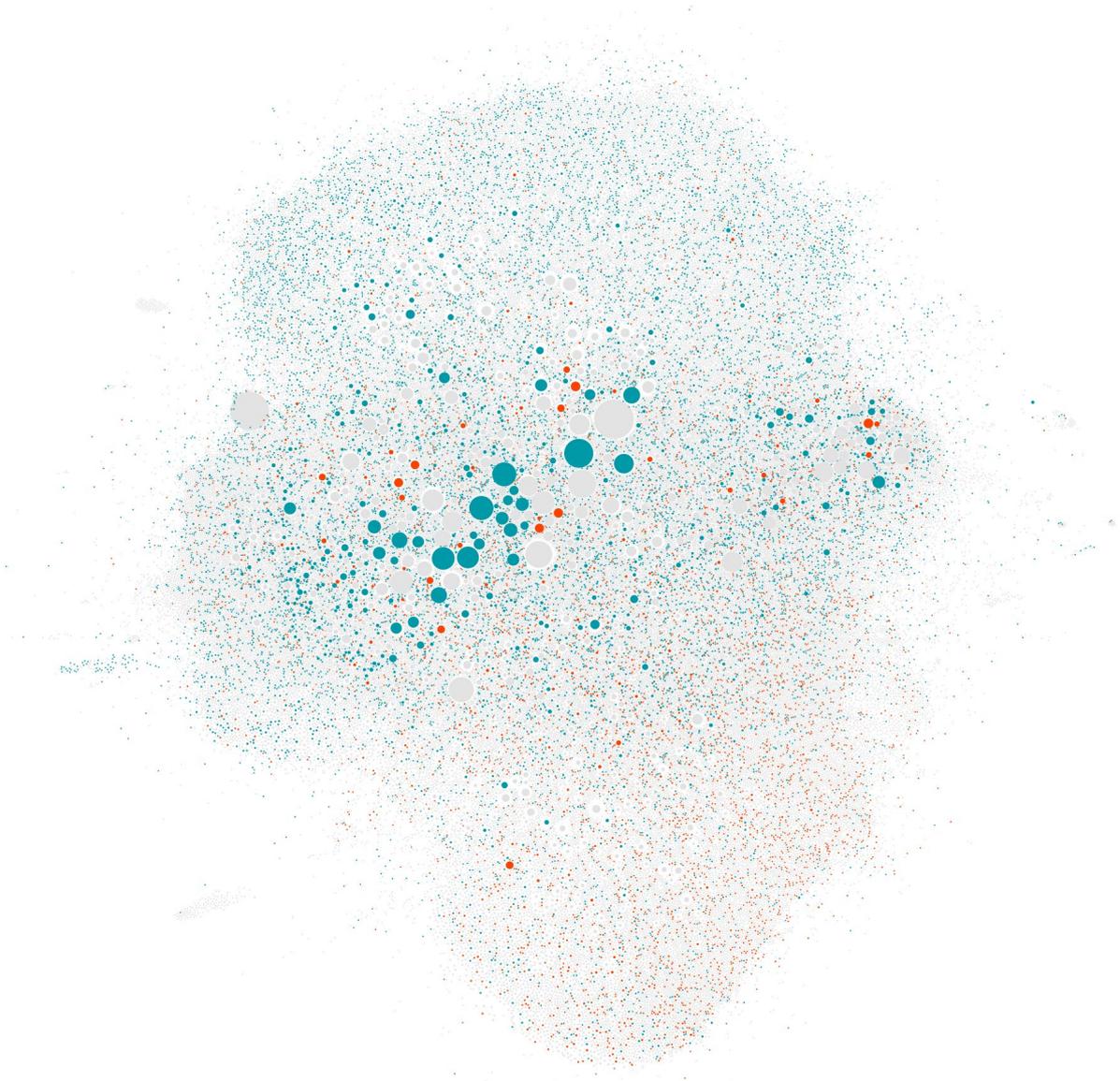

**Figure 7:** A network visualization representing the subset of 98,663 Twitter users who posted tweets including links to vaccine-related webpages at least twice and were connected to at least one other user in the largest connected component. Users who posted at least 2 high credibility webpages and no low credibility webpages (cyan) and users who posted at least 2 low credibility webpages and no high credibility webpages (orange) are highlighted. The size of the nodes is proportional to the number of followers each user has on Twitter, and nodes are positioned by a heuristic such that well-connected groups of users are more likely to be positioned close together in the network diagram.

# Discussion

We found that it is feasible to produce machine learning classifiers to identify vaccine-related webpages of low credibility. Applying a classifier to vaccine-related webpages shared on Twitter between January 2017 and March 2018, we found that fewer low credibility web pages were shared overall, though some had a potential reach of tens of millions of Twitter users. A network visualization suggested that certain communities of Twitter users were much more likely to share and be exposed to low credibility webpages.

This research extends knowledge related to the surveillance of health misinformation on social media. Where much of the prior research has aimed to label individual social media posts or the claims made on social media by veracity (26-30), we instead label webpages shared on social media using a credibility appraisal checklist extended from previously validated instruments to be appropriate to vaccine-related communications (22, 23). In other related work, Mitra et al. (52) examined the linguistic features in social media posts that influence perceptions of credibility. While we did not examine the linguistic features of the tweets that included links to low credibility information, it would be interesting to connect these ideas to better understand whether they influence user behaviour—making users more likely to engage with a tweet by URL access, replying, and sharing.

The work presented here is also different from previous studies examining opinions and attitudes expressed by Twitter users, which mostly label individual tweets or users based on whether they are promoting vaccination or advocating against vaccines (31, 33, 36, 39). Here we consider the communications shared on Twitter rather than the opinions expressed by users in the text of tweets.

Our work is also not directly comparable to previous studies that have examined how misinformation spreads through social media (2-6). We examined a single topic that may not generalise to other application domains such as politics; labelled information according to a broader set of criteria than just the veracity of the information; and measured total potential exposures rather than just cascades of tweets and retweets. Rather than sampling from a set of known examples of fake and real news to compare spread, we sampled from across the spectrum of relevant articles shared on Twitter.

Structuring the experiments in this way, we found no clear difference in the distribution of total potential exposures between low credibility webpages and others. While most low credibility webpages are shared with a smaller number of Twitter users, some had the potential to reach tens of millions.

This work has implications for public health. The ability to measure how people engage with and share misinformation on social media may help us better target and monitor the impact of public health interventions (40-42). We found that certain sub-populations of Twitter users share low credibility vaccine communications more often and are less likely to be connected to users sharing higher credibility vaccine communications. While these results are unsurprising, most studies examining vaccines on social media have only counted tweets rather than examining the heterogeneity of potential exposure to vaccine critical posts (31, 39, 53), despite evidence of the clustering of opinions from as early as 2011 (33). The present study is consistent with these previous findings on clustering, and with studies examining exposure to different topics about HPV vaccines (36, 38). Knowing where low credibility communications are most commonly shared on social media may support the development of communication interventions targeted specifically at communities that are most likely to benefit (54). While the methods are not yet precise enough to reliably identify individual links to low credibility communications, they may eventually be useful as the basis for countermeasures such as active debunking. Methods for inoculating against misinformation by providing warnings immediately prior to access have mixed results (10, 55, 56).

There were several limitations to this study. The first relates to the limited size of the manually labelled sample used for training and internal validation. Our results showed that the DL-based classifiers were less accurate than the RF-based classifiers, but this may have been the consequence of the available training data rather than the general value of the deep learning approach. Without testing on larger sets of training data we are unable to reliably conclude about the comparative performance of the machine learning methods. In addition, other methods and architectures could have been used to predict credibility from text. For example, we could have used simpler methods including Naïve Bayes and logistic regression; used a single multi-label classifier to predict whether a document extracted from a web page satisfied any of the criteria; or constructed a model that directly predicts the credibility score rather than the individual components.

A further limitation relates to the external validity of the classifier. We included only webpages from which we could extract contiguous blocks of text and used a novel approach to sampling from those webpages to create a reasonably balanced sample across the set of credibility scores. Other URLs included in vaccine-related tweets included links to other social media posts (including links to other tweets), links to YouTube and Instagram, links to memes in which text is embedded in an image, links to dynamic pages that no longer show the same information and a range of other pages that included videos or images alongside a small amount of text. Because we were unable to estimate the credibility of the vaccine-related information presented on these other webpages, our conclusions are limited to the characterisation of text-based online webpages. It is likely that a substantial proportion of Instagram, Facebook, and YouTube webpages would receive a low credibility score if they were assessed (57-59), which means we may have underestimated the sharing of low credibility vaccine-related communications on Twitter.

Our estimates of exposure were imperfect. To estimate how many Twitter users might have been exposed to information relative to credibility, we summed the total number of followers of a user for each user that posted the link. We did not count the total number of unique followers who may have seen the link, did not report the number of likes, and do not have access to the number of replies. In the absence of more detailed measures of engagement that can estimate the number of times a webpage was accessed via Twitter, we felt measures of potential exposure were a reasonable upper bound. The conclusions related to measures of potential exposure therefore need to be interpreted with caution and further studies using robust epidemiological designs are needed to reliably estimate exposure.

## Conclusions

We developed and tested machine learning methods to support the automatic credibility appraisal of vaccine-related information online, showing that it is feasible. This allowed us to scale our analysis of large-scale patterns of potential exposure to low credibility vaccine-related webpages shared on Twitter. We found that although low credibility webpages were shared less often overall, there are certain sub-populations where the sharing of low credibility webpages is common. The results suggest two new ways to address the challenge of misinformation, including ongoing surveillance to identify at-risk communities and better target resources in health promotion, and embedding the tool in interventions that flag low credibility communications for consumers as they engage with links to webpages on social media.

## Acknowledgments

Funding: National Health & Medical Research Council (NHMRC) Project Grant APP1128968. Paige Martin contributed to the research with the management of the database, data collection, and data synthesis; and Maryke Steffens contributed to the design and development of the credibility criteria, and the expert labelling.